\documentclass[12pt]{article}

\usepackage{a4}
\begin{document}

\setcounter{page}{1}
\title{Elementary Proof of Moretti's Polar Decomposition Theorem for 
Lorentz Transformations}
\author{H.K. Urbantke\\
{\small Institut f\"ur Theoretische Physik,}
{\small Universit\"at Wien}\\
{\small Boltzmanngasse 5,}
{\small A-1090 Vienna, Austria}}
\date{22 November 2002}
\maketitle
\begin{abstract}

A proof more elementary than the original one is given for Moretti's 
theorem that the usual polar decomposition of real matrices when applied 
to an orthochronous proper Lorentz matrix yields just its standard 
rotation-boost decomposition. (The complex SL(2,C) analog is well-known.)

\end{abstract}


\subsection*{Moretti's polar decomposition theorem for Lorentz matrices 
and 
an elementary proof.}

Recently V. Moretti [1] pointed out that the polar decomposition of real 
matrices when applied to a proper orthochronous Lorentz matrix, 
$L=UP\in{\cal L}_{+}^{\uparrow}$,
where $P$ is symmetric-positive and $U$ is orthogonal, is identical to its 
well-known standard rotation-boost decomposition (Cartan decomposition). 
The apparently non-trivial aspect of this consists in the fact that 
"symmetric-positive" and "orthogonal" refer to the standard real Hilbert 
space structure of ${\bf R^4}$, whereas Lorentz transformations refer to 
Minkowski geometry, so that it is not immediately clear that separately 
$U\in{\cal L}_{+}^{\uparrow}$, $P\in{\cal L}_{+}^{\uparrow}$. But it has 
to be observed that in addition to the Minkowski geometry we here have 
singled out a reference frame $\Sigma$, mathematically given by the 
canonical basis of ${\bf R^4}$, by saying that the Lorentz matrix $L$ 
refers to it. This additional structure allows to relate the Minkowski and 
Hilbert space geometries: if $u$ is the 4-velocity of $\Sigma$, considered 
as a covector using the Minkowski metric $\eta$ (signature $-,+,+,+$), 
then the Hilbert space metric is $\eta +2 u\otimes u$. (Cf. [2, p. 145]; 
for a similar discussion of the geometrical origin of the Hilbert space 
structure used in the analogous polar decomposition of SL(2,C) matrices, 
also treated in [1], see [2, p. 254].)


The elementary proof of Moretti's theorem simply consists in the remark 
that the (unique) standard rotation-boost decomposition 
$L=L_RL_{{\bf v}}$, where

$$L_R:=\left(\begin{array}{cc}1&{\bf 0}^{\top}\\{\bf 
0}&R\end{array}\right),\,\,\, 
L_{{\bf v}}:=\left(\begin{array}{cc}\gamma&-\gamma\,{\bf v}^{\top}\\
-\gamma\,{\bf v}&{\bf 1}+\frac{\gamma^2}{1+\gamma}\,{\bf 
v\,v}^{\top}\end{array}\right)$$
with R$\in$SO(3), ${\bf v}\in{\bf R^3}$, $|{\bf v}|<1$, $\gamma:=(1-{\bf 
v}^2)^{-1/2}$ {\em is} a---and, by the uniqueness of polar decompositions, 
is {\em the}---polar decomposition of $L$: since $L_R\in$SO(4) and the 
symmetry of $L_{\bf v}$ are obvious, it remains to conclude positive 
definiteness for the latter in the case ${\bf v}\neq 0$ from the easily 
checked identity

$$(t\,\,{\bf x}^{\top})L_{\bf v}\left(\begin{array}{c}t\\{\bf 
x}\end{array}\right)
\equiv \gamma(t-{\bf vx})^2+\frac{1}{\gamma}\frac{({\bf vx})^2}{{\bf v}^2}
+({\bf x}-\frac{{\bf vx}}{{\bf v}^2}{\bf v})^2.$$

For completeness, let us recall one way [2, p.11] of deriving the 
rotation-boost decomposition. One first finds the velocity ${\bf v}$ of 
the spatial origin of the inertial frame to which $L$ transforms, starting 
from $\Sigma$: using $L^{-1}=\eta L^{\top}\eta$ one finds $v_i=-L^0_i/L^0_0$. 
One then forms $L\,L^{-1}_{\bf v}$, which checks to be of the form $L_R$, 
where $R$ must be proper-orthogonal from $L\in{\cal L}_{+}^{\uparrow}$, 
$L_{\bf v}\in{\cal L}_{+}^{\uparrow}$.

The polar decomposition $L=P'U'$ with reversed order corresponds to the 
boost-rotation decomposition $L=L_{R{\bf v}}L_R$.

\subsection*{References}
\begin{description}

\item[[1]] V. Moretti, The interplay of the polar decomposition theorem 
and the Lorentz group. arXiv:math-ph/0211047.\\
\enlargethispage{100pt}
\item[[2]] R.U. Sexl, H.K. Urbantke, {\em Relativity, Groups, Particles. 
Special Relativity and Relativistic Symmetry in Field and Particle Physics.} 
Springer, Wien NewYork 2001.

\end{description}

\end{document}